\documentclass[aps,prd,twocolumn,showpacs,superscriptaddress,groupedaddress]{revtex4}  
\usepackage{graphicx}  
\usepackage{dcolumn}   
\usepackage{bm}        
\usepackage{amssymb}   

\begin{document}



\title{(2+1)-dimensional Chern-Simons bi-gravity with AdS Lie bialgebra as an interacting theory of two massless spin-2 fields}
\author { \small{ \bf S. Hoseinzadeh } { \small
and}
\small{ \bf A. Rezaei-Aghdam } \\
{\small{\em Department of Physics, Faculty of Science, Azarbaijan Shahid Madani University, }}\\
{\small{\em 53714-161, Tabriz, Iran }}}

\date{\today}

\begin{abstract}
We introduce a new Lie bialgebra structure for the anti de Sitter (AdS) Lie algebra in (2+1)-dimensional spacetime. By gauging the resulting \textit{AdS Lie bialgebra}, we write a Chern-Simons gauge theory of bi-gravity involving two dreibeins rather than two metrics, which describes two interacting massless spin-2 fields. Our ghost-free bi-gravity model which has no any local degrees of freedom, has also a suitable free field limit.
By solving its equations of motion, we obtain a \textit{new black hole} solution which has two curvature singularities and two horizons. We also study cosmological implications of this massless bi-gravity model.
\end{abstract}

\maketitle

\section{Introduction}

There are different theories of gravity in three-dimensional spacetime and each of them has own advantages and has been widely studied. General relativity is a classical theory which describes interactions of a single massless spin-2 particle (graviton) \cite{1E.Witten,1S.Carlip,2S.Carlip,3S.Carlip,2E.Witten}. Three-dimensional general relativity, without cosmological constant, is equivalent to a Chern-Simons gauge theory with the Poincar{\'e} gauge group ISO(2,1) \cite{1E.Witten}. But, the Chern-Simons gauge theories with gauge groups SO(2,2) or SO(3,1) are equivalent to adding negative or positive cosmological constants to three-dimensional general relativity, respectively \cite{1E.Witten}.

It has been shown that does not exist any consistent theory (with at most two derivatives of the fields) involving interactions of many massless spin-2 fields in spacetime dimensions
$d\!>\!3$, because of the appearance of an unphysical scalar mode of negative energy (Boulware-Deser ghost) in such theories, or their discontinuity in the number of local degrees of freedom at their free field limits \cite{1Boulanger}.
Although, in (2+1)-dimensional spacetime an exotic consistent interacting theory
of many massless spin-2 fields has been constructed in \cite{2Boulanger}, but the physical consequences of such interacting model has not been studied in detail. Theories which describe massless spin-2 fields in (2+1)-dimensional spacetime, have no local degrees of freedom, hence the Chern-Simons theory (with any gauge group) which is a topological model and has no local degrees of freedom \cite{W.Merbis}, is a suitable candidate to construct a (2+1)-dimensional interacting theory of massless spin-2 fields.

On the other hand, in past years, ``massive gravity'' theories which have local degrees of freedom and describe the interactions of the massive spin-2 fields (gravitons), have been developed.
Massive gravity theories have been greatly studied after resolution of their theoretical difficulties (see for a review \cite{1K.Hinterbichler,1C.de Rham}).
Topologically massive gravity \cite{1S.Deser, 2S.Deser, 4S.Carlip, 1M.Blagojevic}, new massive gravity (NMG) \cite{1E.A.Bergshoeff, 2E.A.Bergshoeff, 2M.Blagojevic, O.Hohm} and general massive gravity \cite{O.Hohm} are three higher derivative theories of massive gravity involving auxiliary fields.
dRGT massive gravity \cite{2C.de Rham,3C.de Rham,1S.F.Hassan,W.Merbis} is a bi-metric theory of massive gravity, and describes a massive together with a massless spin-2 particles.
The non-dynamical reference metric of the dRGT model is promoted to a dynamical metric by introducing a kinetic term for it, resulting in the zwei-dreibein gravity (ZDG) \cite{3E.A.Bergshoeff,4E.A.Bergshoeff} (see also \cite{2S.F.Hassan,2K.Hinterbichler}).
The ZDG model has been generalized to obtain a parity-violating model which is called General Zwei-Dreibein Gravity (GZDG) \cite{W.Merbis,5E.A.Bergshoeff}.
These massive gravity models have not Chern-Simons formulations, but they are Chern-Simons-like theories of gravity (see for a review \cite{W.Merbis}).
In ref. \cite{M.R.Setare}, the GZDG$^{+}$ model has been introduced by adding a constraint term to the GZDG model for fixing torsion.
Moreover, during the past few years two different extensions of the Poincar\'{e} algebra, i.e. the Maxwell algebra \cite{H.Bacry,R.Schrader,1J.A.de Azcarraga,2J.A.de Azcarraga,P.Salgado,1O.Cebecioglu,2O.Cebecioglu,1S.Hoseinzadeh} and the semi-simple extension of the Poincar\'{e} algebra (AdS-Lorentz algebra) \cite{1S.Hoseinzadeh,D.V.Soroka,J.Diaz,2S.Hoseinzadeh} have been applied to construct some group-theoretical gravity theories in four and three spacetime dimensions.
Recently, we have studied a (2+1)-dimensional interacting model of two massless spin-2 fields by gauging a new Lie algebra \cite{3S.Hoseinzadeh}. Now, in this paper, we are interested in the study of an interacting theory of two massless spin-2 fields which obtains by gauging a new Lie bialgebra. The resulting bi-gravity model, just like the ZDG model, has been formulated in terms of two dreibeins rather than two metrics. But, unlike the ZDG model, it is a massless zwei-dreibein gravity theory.

Formulating new theories of the gravitational interaction is useful to understand the recent observational data in cosmology, which indicate that the expansion of the universe is accelerating. One of the possibilities for constructing new theories of gravity is extending known classical field theories to include additional spin-2 fields and interactions, which can modify the general relativity at large distances. Bi-metric theory of gravity, which describes the interactions of two different spin-2 fields, is therefore an interesting candidate to explain the accelerated expansion of the universe. One of the motivations of massive and bi-metric theories of gravity is that the interactions could change some of the dynamics of the gravitational theory, and therefore by changing the long-distance behavior of the gravitational fields, make them candidate theories of dark matter and energy.

The outline of the paper is as follows:
In section two, we construct a new \textit{Lie bialgebra} using the AdS Lie algebra so(2,2) in (2+1)-dimensional spacetime.
In Section three, using the obtained Lie bialgebra (and the corresponding Manin triple), we propose a Chern-Simons gauge invariant bi-metric theory of gravity involving two different dreibein fields which describes \textit{two interacting massless spin-2 fields}.
We compare our bi-gravity model with the ``massive gravity'' theories such as NMG and ZDG. We also solve the equations of motion using the BTZ black hole metric for one of the metrics, and obtain a new black hole in the other metric solution.
In section four, we study cosmological implications of the model, and show that it admits a homogeneous and isotropic Friedmann-Robertson-Walker solution.
Some concluding remarks and discussions are given at the end.

\section{Ads Lie bialgebra }

In this section, we introduce a new bialgebra \cite{V.Chari,A.Rezaei-Aghdam} which is obtained by use of the AdS Lie algebra in (2+1)-dimensional spacetime.
In (2+1)-dimensional spacetime, the commutation relations of the six-dimensional AdS Lie algebra are as follows \cite{1E.Witten}:
\begin{eqnarray} \label{AdS Lie algebra}
[J_{a},\! J_{b}] \!=\! \epsilon_{abc} J^{c}\!,~~ [J_{a},\! P_{b}] \!=\! \epsilon_{abc} P^{c}\!,~~ [P_{a},\! P_{b}] \!=\!\frac{1}{\ell^{2}} \epsilon_{abc} J^{c}\!,~
\end{eqnarray}
where $\ell^{-2}=-\Lambda$ is a constant, $\epsilon_{012}=-1$, and $J_{a}$ and $P_{a}$ ($a=0,1,2$) are the Lorentz and translation generators, respectively \footnote{Here, we use $J^{a}=\frac{1}{2}\epsilon^{abc}J_{bc}$ for the Lorentz generators $J_{ab}$.}. The algebra indices $a,b,c$ can be raised and lowered by (2+1)-dimensional Minkowski metric $\eta_{ab}=diag(-1,1,1)$.
By letting the basis of the AdS algebra as $\{X_{1},...,X_{6}\}=\{J_{0},J_{1},J_{2},P_{0},P_{1},P_{2}\}$, and using the structure constants $f_{ab}^{~~c}$ of the AdS Lie algebra (\ref{AdS Lie algebra}) and the following Jacobi and mixed-Jacobi identities \cite{A.Rezaei-Aghdam}:
\begin{eqnarray}\nonumber
\tilde{f}^{ab}_{~~n} \tilde{f}^{nc}_{~~m} \!+\!\tilde{f}^{ca}_{~~n} \tilde{f}^{nb}_{~~m} \!+\!\tilde{f}^{bc}_{~~n} \tilde{f}^{na}_{~~m}\!=\! 0, ~~~~~~~~~~
\nonumber
\\ \nonumber
f_{mc}^{~~~a} \tilde{f}^{bm}_{~~~n}
\!-\! f_{mn}^{~~~a} \tilde{f}^{bm}_{~~~c} \!-\! f_{mc}^{~~~b} \tilde{f}^{am}_{~~~n}
\!+\! f_{mn}^{~~~b} \tilde{f}^{am}_{~~~c} \!=\! f_{cn}^{~~m} \tilde{f}^{ab}_{~~m},
\end{eqnarray}
respectively, we obtain the following structure constants $\tilde{f}^{ab}_{~~c}$ of the dual Lie algebra:
\begin{eqnarray}
\tilde{f}^{54}_{~~6}\!=\!-a,~~~~~ \tilde{f}^{15}_{~~3}\!=\!a,~~~~~ \tilde{f}^{26}_{~~1}\!=\!a,~~~~~ \tilde{f}^{21}_{~~6}\!=\!-\Lambda a,
\nonumber
\\
\tilde{f}^{56}_{~~4}\!=\!-a,~~~~~ \tilde{f}^{35}_{~~1}\!=\!a,~~~~~ \tilde{f}^{24}_{~~3}\!=\!a,~~~~~ \tilde{f}^{23}_{~~4}\!=\!-\Lambda a,
\end{eqnarray}
where ~$a$~ is an arbitrary constant.
By letting the basis of the dual Lie algebra as
$\{\tilde{X}^{1},...,\tilde{X}^{6}\}=\{\tilde{P}_{0},\tilde{P}_{1},\tilde{P}_{2},\tilde{J}_{0},\tilde{J}_{1},\tilde{J}_{2}\}$,
the commutation relations of the dual Lie algebra can be written in the following form:
\begin{eqnarray}
[\tilde{J}_{1},\tilde{J}_{b}] =-a \epsilon_{1bc} \tilde{J}^{c},~~~~~~~~
[\tilde{J}_{b},\tilde{P}_{1}] =-a \epsilon_{1bc} \tilde{P}^{c}, ~~
\nonumber
\\  \label{dual Lie algebra}
[\tilde{J}_{1},\tilde{P}_{b}] =-a \epsilon_{1bc} \tilde{P}^{c},~~~~~~~~
[\tilde{P}_{1},\tilde{P}_{b}] = a\ell^{-2} \epsilon_{1bc} \tilde{J}^{c},
\end{eqnarray}
where $\tilde{J}_{a}$ and $\tilde{P}_{a}$ ($a=0,1,2$) are the generators of spacetime rotation and translation related to the dual geometric structures such as metric and spin connection (see below).
The dual Lie algebra (\ref{dual Lie algebra}) is very similar to the AdS Lie algebra (\ref{AdS Lie algebra}), but in (\ref{dual Lie algebra}) we have the commutation relations between generators with indice "$1$" $(J_{1},P_{1})$ and generators with indice "$j=0,2$" $(J_{j},P_{j})$, only. In other words, generators with indice "$j=0,2$" $(J_{j},P_{j})$ commute with each other. The commutation relations (\ref{AdS Lie algebra}) together with (\ref{dual Lie algebra}) describe \textit{AdS Lie bialgebra}.
Now, using ~$[X_{a},\tilde{X}^{b}]= \tilde{f}^{bc}_{~~~a} X_{c} +f_{ca}^{~~b} \tilde{X}^{c}$ \cite{V.Chari,A.Rezaei-Aghdam}, one can obtain the commutation relations between the generators of the AdS Lie algebra $J_{a},P_{a}$ and the generators of the dual Lie algebra $\tilde{J}_{a},\tilde{P}_{a}$ as follows:
\begin{eqnarray}
[J_{b},\!\tilde{P}_{1}] \!=\!\epsilon_{1bc} (a P^{c} \!-\!\tilde{P}_{c}),~~~
[P_{b},\!\tilde{J}_{1}] \!=\!-\epsilon_{1bc} (a P^{c} \!+\!\tilde{P}_{c}),~
\nonumber
\\  \nonumber
[J_{b},\!\tilde{J}_{1}] \!=\!-\epsilon_{1bc} (a J^{c} \!\!+\!\tilde{J}_{c}),~~
[P_{b},\!\tilde{P}_{1}] \!=\! \ell^{-2} \epsilon_{1bc} (a J^{c} \!\!-\!\tilde{J}_{c}),
\nonumber
\\ \nonumber
[P_{i},\tilde{P}_{j}] \!=\! \ell^{-2} [J_{i},\tilde{J}_{j}] \!=\! \ell^{-2} \epsilon_{1i}^{~~j} (-a J_{1} \!+\!\tilde{J}_{1}),~~~~~~~~~~~~~~~
\nonumber
\\ \nonumber
[J_{i},\tilde{P}_{j}] \!=\! [P_{i},\tilde{J}_{j}] \!=\! \epsilon_{1i}^{~~j} (a P_{1} \!+\!\tilde{P}_{1}),~~~~~~~~~~~~~~~~~~~~~~~~~~
\nonumber
\\  \nonumber
[P_{1},\tilde{P}_{b}] \!=\! \ell^{-2} [J_{1},\tilde{J}_{b}] \!=\!-\ell^{-2} \epsilon_{1bc} \tilde{J}^{c},~~~~~~~~~~~~~~~~~~~~~~~~
\nonumber
\\  \label{mixed commutators}
[J_{1},\tilde{P}_{b}] \!=\! [P_{1},\tilde{J}_{b}] \!=\! -\epsilon_{1bc} \tilde{P}^{c},~~~~~~~~~~~~~~~~~~~~~~~~~~~~~~~~
\end{eqnarray}
where the indices $i$ and $j$ ($i,j=0,2$) are the algebra indices.
The commutation relations (\ref{mixed commutators}) together with (\ref{AdS Lie algebra}) and (\ref{dual Lie algebra}), describe \textit{AdS Manin triple} which is a 12-dimensional Lie algebra.

\section {(2+1)-dimensional Chern-Simons gravity with Ads Lie bialgebra}

In this section, we use the AdS Lie bialgebra, which is discussed in the previous section, to construct a new (2+1)-dimensional Chern-Simons bi-gravity.
Using the relation ~$f_{AB}^{~~~C} ~\Omega_{CD}+f_{AD}^{~~~C}~\Omega_{CB}=0$ \cite{C.Nappi}, \footnote{$f_{AB}^{~~~C}$ is the structure constant of the AdS Manin triple.} an ad-invariant metric $\Omega_{AB}=\langle X_{A},X_{B} \rangle$ for the AdS Manin triple is obtained as follows:
\begin{eqnarray}
\langle J_{a},\! P_{b}\rangle \!=\!-\alpha\!~ \eta_{ab},~~~~~~~~~
\langle J_{a},\!\tilde{P}_{b}\rangle \!=\!\beta \delta_{ab} \!+\! a\alpha~\!\delta_{a1}\delta_{b1},
\nonumber
\\  \nonumber
\langle \tilde{J}_{a},\!\tilde{P}_{b}\rangle \!=\!a^{2}\alpha~\! \delta_{a1}\delta_{b1},~~~~~
\langle P_{a},\!\tilde{J}_{b}\rangle \!=\!\beta \delta_{ab} \!-\! a\alpha~\!\delta_{a1}\delta_{b1},
\nonumber
\\  \nonumber
\langle J_{a},J_{b}\rangle \!=\!\langle P_{a},P_{b}\rangle \!=\!\langle \tilde{J}_{a},\tilde{J}_{b}\rangle \!=\! 0,~~~~~~~~~~~~~~~~~~~~~~~~
\nonumber
\\ \label{ad-invariant metric}
\langle \tilde{P}_{a},\tilde{P}_{b}\rangle \!=\!\langle J_{a},\tilde{J}_{b}\rangle \!=\!\langle P_{a},\tilde{P}_{b}\rangle \!=\! 0,~~~~~~~~~~~~~~~~~~~~~~~~
\end{eqnarray}
where $\delta_{ab}=diag(1,1,1)$ is the Kronecker delta function, and $\alpha$ and $\beta$ are arbitrary constants. The ad-invariant metric should be non-degenerate, and then, we have $\beta\neq0$.
Now, we use the AdS \textit{Manin triple} to construct a gauge symmetric Chern-Simons action,
$I_{cs}=\frac{1}{4\pi}\int_{M} \Big(\langle h \wedge dh \rangle +
\frac{1}{3}~\langle h \wedge [h \wedge h]\rangle \Big),$
where $h=h_{\mu}~dx^{\mu}$ is an AdS Manin triple valued Murer-Cartan one-form gauge field as follows:
\begin{eqnarray}\label{gauge field}
h_{\mu}=h_{\mu}^{~B}X_{B}=e_{\mu}^{~a}P_{a}+\omega_{\mu}^{~a}J_{a}+\tilde{e}_{\mu}^{~a}\tilde{P}_{a}+\tilde{\omega}_{\mu}^{~a}\tilde{J}_{a},
\end{eqnarray}
where the Greek indices $\mu=0,1,2$ are the spacetime indices, $e_{\mu}^{~a}$ and $\omega_{\mu}^{~a}$ are the ordinary dreibein and spin
connection, and $\tilde{e}_{\mu}^{~a} , \tilde{\omega}_{\mu}^{~a}$ are dreibein and spin
connection corresponding to the generators of the dual Lie algebra, respectively. In this point of view, we obtain a gauge theory which has two metric tensors $g_{\mu\nu}=e_{\mu}^{~a} e_{\nu}^{~b} \eta_{ab}$ and $f_{\mu\nu}=\tilde{e}_{\mu}^{~a} \tilde{e}_{\nu}^{~b} \eta_{ab}$. We use the infinitesimal gauge parameter $u=\rho^{a}P_{a}+\tau^{a}J_{a}+\tilde{\rho}^{a}\tilde{P}_{a}+\tilde{\tau}^{a}\tilde{J}_{a}$
together with the commutation relations (\ref{AdS Lie algebra}),(\ref{dual Lie algebra}),(\ref{mixed commutators}) and the gauge transformations $h_{\mu} \rightarrow h^{\prime}_{\mu}\!=\!U^{-1}h_{\mu}U\!+\!U^{-1}\partial_{\mu}U,$ with ~$U\!=\!e^{-u} \simeq  1-u$~ and ~$U^{-1}\!=\!e^{u} \simeq 1+u$,~
to obtain the following transformations for the gauge fields:
\begin{eqnarray}
\delta e_{\mu}^{~c}\!=\!-\partial_{\mu}\rho^{c}\!+\!\epsilon^{abc} (\rho_{a}~\omega_{\mu b} \!+\!\tau_{a}~e_{\mu b}) ~~~~~~~~~~~~~~~~~~~~~~
\nonumber
\\  \nonumber
\!+ a \epsilon^{ibc} \Big( \rho_{i}~\tilde{\omega}_{\mu}^{~b} \!-\!\tilde{\tau}^{b} e_{\mu i}
\!-\!(-1)^{c} (\tau_{i}~\tilde{e}_{\mu}^{~b} \!-\!\tilde{\rho}^{b} \omega_{\mu i}) \Big),
\nonumber
\\  \nonumber
\delta \omega_{\mu}^{~c}\!=\!-\partial_{\mu}\tau^{c}\!+\!\epsilon^{abc} (\ell^{-2} \rho_{a}~ e_{\mu b} \!+\!\tau_{a}~ \omega_{\mu b}) ~~~~~~~~~~~~~~~~~~
\nonumber
\\  \nonumber
\!- a \epsilon^{ibc} \Big(\!\ell^{-2} (\rho_{i}~\!\tilde{e}_{\!\mu}^{\!~b} \!\!-\!\tilde{\rho}^{b} e_{\!\mu i})
\!-\!(-1)^{c} (\tau_{i}~\!\tilde{\omega}_{\!\mu}^{\!~b} \!\!-\!\tilde{\tau}^{b} \omega_{\!\mu i}) \!\Big),
\nonumber
\\  \nonumber
\delta \tilde{e}_{\mu}^{~c}\!=\!-\partial_{\mu}\tilde{\rho}^{c}
\!+\!\epsilon^{abc} \Big( \rho^{b} \tilde{\omega}_{\mu a} \!-\!\tilde{\tau}_{a} e_{\mu}^{~b}
\!+\!\tau^{b} \tilde{e}_{\mu a} \!-\!\tilde{\rho}_{a} \omega_{\mu}^{~b} \Big)~~~~
\nonumber
\\  \nonumber
\!- a \epsilon^{1bc} \Big( \tilde{\tau}_{b}~\tilde{e}_{\mu}^{~1} \!-\!\tilde{\rho}^{1} \tilde{\omega}_{\mu b}
\!+\!\tilde{\tau}^{1} \tilde{e}_{\mu b} \!-\!\tilde{\rho}_{b}~\tilde{\omega}_{\mu}^{~1} \Big),~~~~~~~~
\nonumber
\\  \nonumber
\delta \tilde{\omega}_{\!\mu}^{\!~c}\!\!=\!-\partial_{\mu}\tilde{\tau}^{c}
\!\!+\!\epsilon^{abc} \Big(\! \ell^{-2}(\rho^{b} \tilde{e}_{\mu a} \!\!-\!\tilde{\rho}_{a} e_{\!\mu}^{\!~b})
\!+\!\tau^{b} \tilde{\omega}_{\mu a} \!\!-\!\tilde{\tau}_{a} \omega_{\!\mu}^{\!~b} \!\Big)~
\nonumber
\\  \label{gauge transformations}
\!+ a \epsilon^{1bc} \Big( \tilde{\tau}_{b}~\tilde{\omega}_{\mu}^{~1} \!-\!\tilde{\tau}^{1} \tilde{\omega}_{\mu b}
\!+\! \ell^{-2}(\tilde{\rho}^{1} \tilde{e}_{\mu b} \!-\!\tilde{\rho}_{b} \tilde{e}_{\mu}^{~1}) \Big).~~~
\end{eqnarray}
The Ricci curvature two-form $\mathcal{R}=\mathcal{R}_{\mu\nu} dx^{\mu} \wedge dx^{\nu}$ can be written as:
\begin{eqnarray}
\mathcal{R}_{\mu\nu}=\partial_{[\mu}h_{\nu]}+[h_{\mu},h_{\nu}] =
\mathcal{R}_{\mu\nu}^{~A} X_{A}~~~~~~~~~~~~~~~~~
\nonumber
\\
=T_{\mu\nu}^{~~a}~P_{a}+R_{\mu\nu}^{~~a}~J_{a}+\tilde{T}_{\mu\nu}^{~~a}~\tilde{P}_{a}+\tilde{R}_{\mu\nu}^{~~a}~\tilde{J}_{a},
\end{eqnarray}
such that the torsion $T_{\mu\nu}^{~~a}$ and the standard Riemannian curvature $R_{\mu\nu}^{~~a}$ are as follows:
\begin{eqnarray}
T_{\mu\nu}^{~~j}=\partial_{[\mu} e_{\nu]}^{~j}
\!+\!\epsilon_{ab}^{~~j}\omega_{[\mu}^{~a} e_{\nu]}^{~b}
\!+\! a\epsilon_{1b}^{~~j} \Big(\omega_{[\mu}^{~b} \tilde{e}_{\nu]}^{~1}
\!-\! e_{[\mu}^{~b} \tilde{\omega}_{\nu]}^{~1}\Big),
\nonumber
\\ \nonumber
T_{\mu\nu}^{~~1}=\partial_{[\mu} e_{\nu]}^{~1}
\!+\!\epsilon_{ab}^{~~1}\omega_{[\mu}^{~a} e_{\nu]}^{~b}
\!+\! a\epsilon_{1a}^{~~b} \Big(\omega_{[\mu}^{~a} \tilde{e}_{\nu]}^{~b}
\!+\! e_{[\mu}^{~a} \tilde{\omega}_{\nu]}^{~b}\Big),
\nonumber
\\ \nonumber
R_{\mu\nu}^{~~j}=\partial_{[\mu} \omega_{\nu]}^{~j} +\frac{1}{2}\epsilon_{ab}^{~~j} \Big( \omega_{[\mu}^{~a} \omega_{\nu]}^{\ b} +\ell^{-2} e_{[\mu}^{~a} e_{\nu]}^{\ b} \Big)~~~~~~~~~
\nonumber
\\ \nonumber
+ a\epsilon_{1b}^{~~j} \Big( \ell^{-2} e_{[\mu}^{~b} \tilde{e}_{\nu]}^{~1}
-\omega_{[\mu}^{~b} \tilde{\omega}_{\nu]}^{~1} \Big),~~~~~~~~~~~~~~~~~~~
\nonumber
\\ \nonumber
R_{\mu\nu}^{~~1}=\partial_{[\mu} \omega_{\nu]}^{~1} +\frac{1}{2}\epsilon_{1ab} \Big( \omega_{[\mu}^{~a} \omega_{\nu]}^{\ b} +\ell^{-2} e_{[\mu}^{~a} e_{\nu]}^{\ b} \Big)~~~~~~~~~
\nonumber
\\ \label{field strength}
- a\epsilon_{1a}^{~~b} \Big( \ell^{-2} e_{[\mu}^{~a} \tilde{e}_{\nu]}^{~b}
+\omega_{[\mu}^{~a} \tilde{\omega}_{\nu]}^{~b} \Big),~~~~~~~~~~~~~~~~~~~
\end{eqnarray}
and in the same way, the field strengths $\tilde{T}_{\mu\nu}^{~~a}$ and $\tilde{R}_{\mu\nu}^{~~a}$ which can be interpreted as dual torsion and dual Riemannian curvature respectively, have the following forms:
\begin{eqnarray}
\tilde{T}_{\mu\nu}^{~~j}\!=\!\partial_{[\mu} \tilde{e}_{\nu]}^{~j}
\!-\!\epsilon_{1bj} \Big(e_{[\mu}^{~b} \tilde{\omega}_{\nu]}^{~1}
\!+\!\omega_{[\mu}^{~b} \tilde{e}_{\nu]}^{~1}\Big)~~~~~~~~~~~~~~~~~~~~~~
\nonumber
\\ \nonumber
\!+ \epsilon_{1b}^{~~j}\Big(a \tilde{e}_{[\mu}^{~1} \tilde{\omega}_{\nu]}^{~b}
\!-\! a \tilde{\omega}_{[\mu}^{~1} \tilde{e}_{\nu]}^{~b}
\!-\! e_{[\mu}^{~1} \tilde{\omega}_{\nu]}^{~b}
\!-\! \omega_{[\mu}^{~1} \tilde{e}_{\nu]}^{~b}\Big),~~~~~~~
\nonumber
\\ \nonumber
\tilde{T}_{\mu\nu}^{~~1}\!=\!\partial_{[\mu}
\tilde{e}_{\nu]}^{~1}
\!+\!\epsilon_{1a}^{~~b} \omega_{[\mu}^{~a} \tilde{e}_{\nu]}^{~b}
\!+\!\epsilon_{1a}^{~~b} e_{[\mu}^{~a} \tilde{\omega}_{\nu]}^{~b},~~~~~~~~~~~~~~~~~~~~
\nonumber
\\ \nonumber
\tilde{R}_{\mu\nu}^{~~j}\!=\!\partial_{[\mu} \tilde{\omega}_{\nu]}^{~j}
\!-\!\epsilon_{1bj}\Big(\ell^{-2} e_{[\mu}^{~b} \tilde{e}_{\nu]}^{~1}
\!+\!\omega_{[\mu}^{~b} \tilde{\omega}_{\nu]}^{~1}\Big)~~~~~~~~~~~~~~~~~
\nonumber
\\ \nonumber
\!-\epsilon_{1b}^{~~j}\Big(\! \ell^{-2} e_{[\mu}^{~1} \tilde{e}_{\nu]}^{~b}
\!+\! a \tilde{\omega}_{[\mu}^{~1} \tilde{\omega}_{\nu]}^{~b}
\!-\!\ell^{-2} a \tilde{e}_{[\mu}^{~1} \tilde{e}_{\nu]}^{~b}
\!+\! \omega_{[\mu}^{~1} \tilde{\omega}_{\nu]}^{~b}\!\Big),
\nonumber
\\  \label{dual field strength}
\tilde{R}_{\mu\nu}^{~~1}\!=\!\partial_{[\mu}
\tilde{\omega}_{\nu]}^{~1} \!+\!\epsilon_{1a}^{~~b} \omega_{[\mu}^{~a} \tilde{\omega}_{\nu]}^{~b} \!+\!\ell^{-2}\epsilon_{1a}^{~~b} e_{[\mu}^{~a} \tilde{e}_{\nu]}^{~b}.~~~~~~~~~~~~~~~
\end{eqnarray}
Using (\ref{AdS Lie algebra}) as well as (\ref{dual Lie algebra})-(\ref{gauge field}), one obtains the following Chern-Simons bi-gravity model with the AdS Manin triple as a gauge symmetry:
\begin{eqnarray}\label{model}
I={I}^{e,\omega}(e,\omega)+{I}^{\tilde{e},\tilde{\omega}}(\tilde{e},\tilde{\omega})+I^{int}(e,\omega,\tilde{e},\tilde{\omega}),
\end{eqnarray}
where the first term is
\begin{eqnarray}\nonumber
{I}^{e,\omega}\!=\!-4\hat{\alpha} G ~I_{\!EC}(\omega,\!e,\!\Lambda),~~~~
\end{eqnarray}
and the Eistein-Cartan action $I_{EC}$ is
\begin{eqnarray}\nonumber
I_{\!EC}(\!\omega,e,\Lambda\!)\!=\!-\frac{1}{16\pi G}\!\!\int_{\!M}\!\!\!\! d^{3}\!x \epsilon^{\mu\nu\rho}
e_{\!\mu}^{\!~c}\!\Big(\! D_{\nu} \omega_{\!\rho c}
\!\!-\!\frac{\Lambda}{3}\epsilon_{abc} e_{\nu}^{~a}
e_{\!\rho}^{\!~b} \!\Big)\!.~
\end{eqnarray}
The second term in (\ref{model}) is the Einstein-Cartan action with $\tilde{\omega}_{\mu}^{~j}\!=\!\tilde{e}_{\mu}^{~j}\!=\!0,~\tilde{\omega}_{\mu}^{~1}\neq 0,~\tilde{e}_{\mu}^{~1}\neq 0,$ as follows:
\begin{eqnarray}\nonumber
I^{\tilde{e},\tilde{\omega}}\!=\!-\frac{\hat{\alpha} a^{2}\!\!}{4\pi} \!\int_{\!M}\!\! d^{3}\!x ~\epsilon^{\mu\nu\rho} \tilde{e}_{\mu}^{~1} \partial_{[\nu} \tilde{\omega}_{\rho]}^{~1}\!.
\end{eqnarray}
The third term in (\ref{model}) includes some interaction terms between the fields $\{e_{\mu}^{~a}, \omega_{\mu}^{~a}\}$ and  the fields $\{\tilde{e}_{\mu}^{~a} , \tilde{\omega}_{\mu}^{~a}\}$ as follows:
\begin{eqnarray}\nonumber
I^{int}\!\!=\!\! \!\int_{\!M}\!\! \frac{d^{3}x}{4\pi} \epsilon^{\mu\nu\rho} \!\Big\{\!a\hat{\alpha} \!\Big(\!\tilde{\omega}_{\mu}^{~1} D_{\!\nu} e_{\!\rho}^{\!~1} \!-\!\tilde{e}_{\mu}^{~1} (D_{\!\nu} \omega_{\!\rho}^{\!~1} \!+\!\ell^{-2} \epsilon_{1bc} e_{\nu}^{~b}\! e_{\rho}^{~c}) \!\Big)
\nonumber
\\ \nonumber
+\hat{\beta} \Big(\!\tilde{\omega}_{\mu}^{~c} D_{\!\nu} e_{\!\rho}^{\!~c}
\!+\!\tilde{e}_{\mu}^{~c} (D_{\!\nu} \omega_{\!\rho}^{\!~c} \!+\!\ell^{-2} \epsilon_{ab}^{~~c} e_{\nu}^{~a} e_{\rho}^{~b}) \!\Big)
\nonumber
\\ \nonumber
\!+2a\hat{\beta} \epsilon_{1b}^{~~c}\! \Big(\! \tilde{\omega}_{\mu}^{~1} (\!\omega_{\nu}^{~b} \tilde{e}_{\rho}^{~c} +e_{\nu}^{~b} \tilde{\omega}_{\rho}^{~c}\!)
\!-\! \tilde{e}_{\mu}^{~1} (\!\omega_{\nu}^{~b} \tilde{\omega}_{\rho}^{~c} \!-\!\ell^{-2} \tilde{e}_{\nu}^{~b} e_{\rho}^{~c}\!) \!\Big)\! \Big\},
\end{eqnarray}
where $D_{\nu} \omega_{\rho}^{~c}$ and $D_{\nu} e_{\rho}^{~c}$ are the covariant derivatives with respect to the spin connection $\omega_{\mu}^{~c}$ as follows:
\begin{eqnarray}
D_{\nu} \omega_{\rho c}=\partial_{[\nu}~\omega_{\rho]c}+
\epsilon_{abc}~\omega_{\nu}^{\ a} \omega_{\rho}^{\ b},
\nonumber
\\
D_{\nu} e_{\rho c}=\partial_{[\nu}~e_{\rho]c}+
\epsilon_{abc}~\omega_{[\nu}^{\ a} e_{\rho]}^{\ b}.~~
\end{eqnarray}
The Chern-Simons bi-gravity model (\ref{model}) which is invariant under the gauge transformations (\ref{gauge transformations}), has no any local degrees of freedom, and is a ghost-free model which describes two interacting massless spin-2 fields in (2+1)-dimensional spacetime.
Two dreibeins in the action (\ref{model}) are related to their corresponding metric tensors as follows:
\begin{eqnarray}
g_{\mu\nu}\!=e_{\mu}^{~a} e_{\nu}^{~b} \eta_{ab},~~~~~~ f_{\mu\nu}\!=\tilde{e}_{\mu}^{~a} \tilde{e}_{\nu}^{~b} \eta_{ab}.
\end{eqnarray}
In the absence of the interaction terms $I^{int}$, the free field limit of (\ref{model}),
\begin{eqnarray}\label{free action}
I^{free}={I}^{e,\omega}(e,\omega) +{I}^{\tilde{e},\tilde{\omega}}(\tilde{e},\tilde{\omega}),
\end{eqnarray}
similar to (\ref{model}) has no any local degrees of freedom, and is invariant under the following gauge transformations:
\begin{eqnarray}
\delta e_{\mu}^{~c}\!=\!-\partial_{\mu}\rho^{c}\!+\!\epsilon^{abc} (\rho_{a} \omega_{\mu b} \!+\!\tau_{a} e_{\mu b}),~~~~~~
\delta \tilde{e}_{\mu}^{~c}\!=\!-\partial_{\mu}\tilde{\rho}^{c}\!,~
\nonumber
\\
\delta \omega_{\!\mu}^{\!~c}\!=\!-\partial_{\mu}\tau^{c}\!\!+\!\epsilon^{abc} (\!\ell^{-2} \rho_{a}  e_{\mu b} \!+\!\tau_{a}  \omega_{\mu b}\!),~~~
\delta \tilde{\omega}_{\!\mu}^{\!~c}\!=\!-\partial_{\mu}\tilde{\tau}^{c}\!.~
\end{eqnarray}
Now, by assuming the following relations among the fields and constants:
\begin{eqnarray}
\beta\!~ \tilde{e}_{\mu}^{~a}\!=\!-\frac{1}{m^{2}}f_{\mu a},~~~~~~~~~~
\beta\!~ \tilde{\omega}_{\mu}^{~a}\!=\! h_{\mu a},~~
\nonumber
\\
\alpha\!=\!-\sigma ,~~~~
\alpha\ell^{-2}\!=\!\Lambda_{0},~~~~
a\ell^{-2}\!=\! 2 m^{2}\beta,
\end{eqnarray}
the Chern-Simons action (\ref{model}) can be rewritten in the following form:
\begin{eqnarray}\nonumber
I=\frac{1}{2\pi} I_{NMG} +\frac{1}{4\pi} \!\!\int\!\! d^{3}\!x  \epsilon^{\mu\nu\rho} \Big\{\!
\!-\! a^{2} \alpha \tilde{e}_{\mu}^{~1} \partial_{[\nu} \tilde{\omega}_{\rho]}^{~1}~~~~~
\nonumber
\\  \nonumber
+ a\alpha \Big(\tilde{\omega}_{\mu}^{~1} D_{\nu} e_{\rho}^{~1} \!-\!\tilde{e}_{\mu}^{~1} (D_{\nu} \omega_{\rho}^{~1}\!+\!\ell^{-2} \epsilon_{1bc} e_{\nu}^{~b} e_{\rho}^{~c}) \Big)~
\nonumber
\\  \nonumber
-a\beta\ell^{-2} \epsilon_{1bc} e_{\mu}^{~1} \tilde{e}_{\nu}^{~b} \tilde{e}_{\rho}^{~c} \!+\!\beta\ell^{-2} \epsilon_{ab}^{~~c} \tilde{e}_{\mu}^{~c} e_{\nu}^{~a} e_{\rho}^{~b}~~~~~~~~
\nonumber
\\  \label{First model-2}
+2a\beta~\epsilon_{1b}^{~~c} \Big( \tilde{\omega}_{\mu}^{~1} (\omega_{\nu}^{~b} \tilde{e}_{\rho}^{~c} \!+\! e_{\nu}^{~b} \tilde{\omega}_{\rho}^{~c}) \!-\!\tilde{e}_{\mu}^{~1} \omega_{\nu}^{~b} \tilde{\omega}_{\rho}^{~c} \Big)
\!\Big\},
\end{eqnarray}
where $I_{NMG}$ is the NMG model involving a pair of the auxiliary fields $f_{\mu}^{~c}, h_{\mu}^{~c}$, as follows: \cite{1E.A.Bergshoeff}
\begin{eqnarray}\nonumber
I_{N\!M\!G}\!=\!\frac{1}{2}\!\!\int\!\! d^{3}\!x \epsilon^{\mu\nu\rho} \Big\{\! \!-\!
\sigma e_{\mu}^{~c} D_{\nu} \omega_{\rho c} \!+\!\frac{\Lambda_{0}}{3}
\epsilon_{abc} e_{\mu}^{~a} e_{\nu}^{~b} e_{\rho}^{~c} ~~~~
\nonumber
\\
+h_{\!\mu}^{\!~c} D_{\nu} e_{\rho c} \!-\!\frac{1}{m^{2}}
f_{\!\mu}^{\!~c} \Big(\! D_{\nu} \omega_{\rho c} \!+\!\epsilon_{abc}
(e_{\!\nu}^{\!~a} f_{\!\rho}^{\!~b} \!+\! f_{\!\nu}^{\!~a} e_{\!\rho}^{\!~b}) \!\Big)
\!\Big\}.~
\end{eqnarray}
Again, using the following redefinition of the fields and constants,
\begin{eqnarray}\nonumber
e_{1}^{\!~a}\equiv e^{a},~~~~~~ \omega_{1}^{\!~a}\equiv \omega^{a},~~~~~~
e_{2}^{\!~a}\equiv \tilde{e}^{a},~~~~~~ \omega_{2}^{\!~a}\equiv \tilde{\omega}^{a},
\nonumber
\\  \nonumber
a^2\!=\! 1,~~~ \alpha\!=\! M_{\!P},~~~ \ell^{-2}\!\!=\!-\alpha_{1} m^{2},~~~ \beta\!=\! M_{\!P}(a\!-\!\frac{\beta_{1}}{\alpha_{1}}),
\end{eqnarray}
the Chern-Simons model (\ref{model}) can be rewritten in another form as follows:
\begin{eqnarray} \nonumber
I=\frac{1}{2\pi} I_{ZDG}(\sigma\!=\!-1, \tilde{e}_{\mu}^{~j}\!=\!\tilde{\omega}_{\mu}^{~j}\!=\!0) ~~~~~~~~~~~~~~~~~~~~~~~~~~~~
\nonumber
\\  \nonumber
+\frac{1}{4\pi} \!\!\int_{\!M}\!\!\! d^{3}\!x  \epsilon^{\mu\nu\rho} \Big\{ \!
a\alpha \Big(\tilde{\omega}_{\mu}^{~1} D_{\nu} e_{\rho}^{~1} \!-\!\tilde{e}_{\mu}^{~1} D_{\nu} \omega_{\rho}^{~1} \Big)~~~~~~~~~~~~~~~
\nonumber
\\  \nonumber
+\beta \Big(\tilde{\omega}_{\mu}^{~c} D_{\nu} e_{\rho}^{~c}\!+\!\tilde{e}_{\mu}^{~c} D_{\nu} \omega_{\rho}^{~c} \!+\!\ell^{-2} \epsilon_{ab}^{~~j} e_{\mu}^{~a} e_{\nu}^{~b}\tilde{e}_{\rho}^{~j} \Big)~~~~~~~~~~~~
\nonumber
\\  \nonumber
+2a\beta \epsilon_{1b}^{~~c} \!\Big(\! \tilde{\omega}_{\!\mu}^{\!~1} (\omega_{\nu}^{~b} \tilde{e}_{\!\rho}^{\!~c} \!+\! e_{\nu}^{~b} \tilde{\omega}_{\!\rho}^{\!~c})
\!- \!\tilde{e}_{\!\mu}^{\!~1} (\omega_{\nu}^{~b} \tilde{\omega}_{\!\rho}^{\!~c} \!-\!\ell^{-2} \tilde{e}_{\nu}^{~b} e_{\!\rho}^{\!~c}) \!\Big)\! \!\Big\}\!,
\end{eqnarray}
where $I_{ZDG}(\sigma\!=\!-1, \tilde{e}_{\mu}^{~j}\!=\!\tilde{\omega}_{\mu}^{~j}\!=\!0)$ is the ZDG action \cite{3E.A.Bergshoeff}:
\begin{eqnarray}\nonumber
I_{ZDG}\!=\!-\frac{1}{2} M_{p} \!\!\int\!\! d^{3}\!x
\epsilon^{\mu\nu\rho} \Big\{ \sigma e_{1\mu}^{~~c} D_{\nu}
\omega_{1\rho c} \!+\!e_{2\mu}^{~~c} D_{\nu} \omega_{2\rho c}~~~~
\nonumber
\\  \nonumber
+\frac{1}{3} \alpha_{1} m^{2} \epsilon_{abc} e_{1\mu}^{~~a}
e_{1\nu}^{~~b} e_{1\rho}^{~~c} \!+\!\frac{1}{3} \alpha_{2} m^{2}
\epsilon_{abc} e_{2\mu}^{~~a} e_{2\nu}^{~~b} e_{2\rho}^{~~c}
\nonumber
\\
-\beta_{1} m^{2} \epsilon_{abc} e_{1\mu}^{~~a} e_{1\nu}^{~~b}
e_{2\rho}^{~~c} \!-\!\beta_{2} m^{2} \epsilon_{abc} e_{1\mu}^{~~a}
e_{2\nu}^{~~b} e_{2\rho}^{~~c} \Big\}\!,~~
\end{eqnarray}
with the sign parameter $\sigma\!=\!-1$ and the fields $\tilde{e}_{\!\mu}^{\!~j}\!=\!\tilde{\omega}_{\!\mu}^{\!~j}\!=\!0,~(j\!=\!0,2)$,
where $M_{p}$ is the Planck mass, $\alpha_{1}$ and $\alpha_{2}$
are cosmological parameters, $\beta_{1}$ and $\beta_{2}$ are
coupling constants, and $e_{I\mu}^{~~a}$ and $\omega_{I\mu}^{~~a}$
($I=1,2$) are pairs of the dreibein and spin connection one-forms,
respectively.
Note that the zero values of the fields $\tilde{e}_{\mu}^{~j}=\tilde{\omega}_{\mu}^{~j}=0,$ in ZDG action is imposed by the dual Lie algebra (\ref{dual Lie algebra}).
Variations of the action (\ref{model}) with respect to the fields $e_{\mu a}, \tilde{e}_{\mu a}, \omega_{\mu a}$ and $\tilde{\omega}_{\mu a}$ give the corresponding equations of motion, respectively:
\begin{eqnarray}\label{e.o.m.}
T_{\nu\rho}^{~~a}=\tilde{T}_{\nu\rho}^{~~a}=R_{\nu\rho}^{~~a}=\tilde{R}_{\nu\rho}^{~~a}=0,
\end{eqnarray}
where $T_{\nu\rho}^{~~a}, \tilde{T}_{\nu\rho}^{~~a}, R_{\nu\rho}^{~~a}$ and $\tilde{R}_{\nu\rho}^{~~a}$ are defined in (\ref{field strength})-(\ref{dual field strength}).

\subsection{Black hole solution}

Now, we use the BTZ black hole metric $g_{\mu\nu}$ \cite{1Banados} to obtain the following solution for the equations of motion (\ref{e.o.m.}):
\begin{eqnarray}\label{BTZ metric}
ds^{2}\!=\!-N^2\!(r) dt^2\!+\! \frac{dr^2}{N^{2}\!(r)}  \!+\! r^2 (\! N^{\varphi}\!(r) dt \!+\! d\varphi \!)^{\!2}\!\!,~~~~~~~~
\\ \label{New metric}
df^{2}\!\!=\!-\frac{\! 4N_{\! f}^4\!\!\!}{\! a^2\! N^{2}} dt^{2}\!\!+\! \frac{\!\! 4(\!\Lambda_{\! f}\! N^{2}\!\!-\!\Lambda N_{\! f}^{2})^{\!2}\!\!}{a^{2}\Lambda^{2} N^{6}} dr^{2}
\!\!+\! r^{2} (\! N_{\! f}^{\varphi}\! dt \!+\! d\varphi \!)^{\!2}\!\!,
\end{eqnarray}
where $ds^{2}=g_{\mu\nu}dx^{\mu}dx^{\nu}$,~~$df^{2}=f_{\mu\nu}dx^{\mu}dx^{\nu}$, and
\begin{eqnarray}\nonumber
N^2(r) \!=\! -M \!+\!\frac{r^{2}}{\ell^{2}}\!+\! \frac{J^{2}}{4 r^{2}} ,~~~~~ N^{\varphi}(r) \!=\!- \frac{J}{2 r^{2}},~~~~~~~~
\nonumber
\\  \nonumber
N_{\!f}^{2}(r)\!=\!-M_{\!f} \!+\!\frac{r^{2}}{\ell_{\!f}^{2}} \!+\!\frac{J^{2}}{4 r^{2}},~~~~ N_{\!f}^{\varphi}(r)\!=\!-\frac{J\!+\! 2D r^{2}}{a r^{2}},~
\end{eqnarray}
$\{x^{0},x^{1},x^{2}\}=\{t,r,\varphi\}$ are the coordinates of the spacetime, $M,J,D,M_{f}$ and $\ell_{\!f}$ are arbitrary constants, and the spin connections $\omega_{\mu}^{~a}(r)$ and $\tilde{\omega}_{\mu}^{~a}(r)$ are obtained as follows:
\begin{eqnarray}\nonumber
{\tiny \omega^{0}\!\!=\! 2D \! N dt \!+\!(\!1\!-\! a\!)\! N d\varphi,} ~~~~~~~~~~~~~~~~~~~~~~~~~~~~~~~~~~~~~~
\nonumber
\\ \nonumber
\omega^{1}\!\!=\!\!\Big(\! \frac{2J\! D}{r}(\!2\!-\!\frac{\ell^{2}}{\ell_{\!f}^{2}}\!) \!+\!(\!\ell^{-2}\!+\! 2\ell_{\!f}^{-2}\!)r\!\Big) dt ~~~~~~~~~~~~~~~~~~~~~~~~~
\nonumber
\\ \nonumber
{\tiny \!+\frac{J}{2 r}\!\Big(\! 2(\!1\!-\! a\!)(\!2\!-\!\frac{\ell^{2}}{\ell_{\!f}^{2}}\!)\!-\!1 \!\Big)\! d\varphi,}~~~~~~~~~~~~~~~~~~~~~~~~~~~~~~
\nonumber
\\ \nonumber
\omega^{2}\!=\!-\frac{J}{2r^{2}\! N} dr, ~~~~~~~~~~~~~~~~~~~~~~~~~~~~~~~~~~~~~~~~~~~~~~~~~
\nonumber
\\ \nonumber
\tilde{\omega}^{0}\!=\!{\tiny \frac{2D}{\!a N\!}}\Big(\!2 N_{\!f}^{2}\!+\! N^{2}(\!1\!-\! 2\frac{\ell^{2}}{\ell_{\!f}^{2}}\!)\!\Big) dt ~~~~~~~~~~~~~~~~~~~~~~~~~~~~~~
\nonumber
\\ \nonumber
+N \Big\{\!\frac{2(\!1\!-\!a\!)}{a}\!\Big(\!\frac{N_{\!f}^{2}}{ N^{2}} \!-\!\frac{\ell^{2}}{\ell_{\!f}^{2}}\!\Big)\!-\!1 \!\Big\} d\varphi, ~~~~~~~~~~~~~~~~~~~~~~~~
\nonumber
\\ \nonumber
\tilde{\omega}^{1}\!\!=\! \Big(\!\frac{J\!D}{a r}(\frac{2\ell^{2}}{\ell_{\!f}^{2}}\!-\!3\!) \!-\!\frac{2r\!}{a\ell_{\!f}^{2}}\!\Big) dt
\!+\!\frac{J}{2 r}\Big(\!\frac{2(\!a\!-\!1\!)}{a}\!(\!1\!-\!\frac{\ell^{2}}{\ell_{\!f}^{2}}\!)\!+\!1 \!\Big) d\varphi,\!\!
\nonumber
\\ \label{AdS spin connection}
\tilde{\omega}^{2}\!=\! \frac{J}{a r^{2}\! N^{3}} \!\Big(\! N^{2} \!-\! N_{\!f}^{2}\!\Big)\! dr.~~~~~~~~~~~~~~~~~~~~~~~~~~~~~~~~~~~~~~
\end{eqnarray}
The Kretschmann scalar $K\!=\!R_{\mu\nu\rho\sigma} R^{\mu\nu\rho\sigma}$ for the metric $f_{\mu\nu}$ is proportional to $N_{\!f}^{-8}\!\!$, and then $f_{\mu\nu}$ has two curvature singularities at
\begin{eqnarray}
{r_{\!s}}_{\pm}\!=\!\sqrt{\frac{\ell_{\!f}}{2}\Big(\ell_{\!f}\!M_{\!f}\!\pm\!\sqrt{\ell_{\!f}^{2}\!M_{\!f}^{2}\!-\! J^{2}}\Big)},~~~~~~
|\ell_{\!f} M_{\!f}|\!>\! |J|,~~
\end{eqnarray}
where $N_{\!f}\!(r)$ vanishes.
$f_{\mu\nu}$ has also two horizons at
\begin{eqnarray}
r_{\!\pm}\!=\!\sqrt{\frac{\ell}{2}\Big(\ell M\!\pm\!\sqrt{\ell^{2}\!M^{2}\!-\! J^{2}}\Big)},~~~~~~~~
|\ell M|\!>\! |J|,~~
\end{eqnarray}
where $N\!(r)$ vanishes.
We use suitable values of the arbitrary constants $M,M_{\!f},\ell$ and $\ell_{\!f}$ to have $r_{\!+}\!\!>\!\!{r_{\!s}}_{+}$, such that $r_{\!+}$ is the event horizon of the black hole.
Then, depending on the values of these constants, we have three different situations:
$r_{\!-}\!\!>\!{r_{\!s}}_{+}$,~ ${r_{\!s}}_{+}\!\!>\!r_{\!-}\!>\!{r_{\!s}}_{-}$ and $r_{\!-}\!\!<\!{r_{\!s}}_{-}$.

To investigate the asymptotic behavior of this solution, we keep only the dominant terms. For very large values of $r$, $ds^{2}$ has the following form:
\begin{eqnarray}\nonumber
ds^{2}~\sim~ -\frac{r^{2}}{\ell^{2}}dt^{2} +\frac{\ell^{2}}{r^{2}}dr^{2} +r^{2}d\varphi^{2},
\end{eqnarray}
which is the AdS spacetime.
But The metric $df^{2}$, for large values of $r$, approaches to the following one:
\begin{eqnarray}\nonumber
df^{2}\! \sim \!-\frac{4\ell^{2} r^{2}\!}{a^{2}\ell_{\!f}^{4}} dt^{2}
\!\!+\!\frac{4\ell^{10}\!{\Big(\!M\!/\!\ell_{\!f}^{2}\!-\! M_{\!f}\!/\!\ell^{2}\!\Big)}^{\!2}\!\!}{a^{2}r^{6}} dr^{2}
\!\!+\!\Big(\!\frac{2Dr\!}{a} dt \!-\!rd\varphi \!\Big)^{\!2}\!\!,~
\end{eqnarray}
which is clearly different from the AdS spacetime.
This \textit{new black hole} is different from the black hole solutions of the three dimensional $f\!\!-\!\!g$ theory, which are asymptotically AdS and have coordinate singularities \cite{2Banados,Afshar}.

\section{Cosmological implications}

We study the homogeneous and isotropic cosmology of our massless bi-gravity model
(\ref{model}) using the following Friedmann-Robertson-Walker (FRW) Ansatz for both metrics:
\begin{eqnarray}\label{FRW-1}
ds^{2}=-N^{2}(t) ~dt^{2}+A^{2}(t) \Big(\frac{dr^{2}}{1\!-\! k r^2}+r^{2}d\varphi^{2}\Big),
\end{eqnarray}
and
\begin{eqnarray}\label{FRW-2}
df^{2}=-X^{2}(t) ~dt^{2}+Y^{2}(t) \Big(\frac{dr^{2}}{1\!-\! k r^2}+r^{2}d\varphi^{2}\Big),
\end{eqnarray}
where $A(t)$ and $Y(t)$ are the spatial scale factors of the FRW metrics $g_{\mu\nu}$ and $f_{\mu\nu}$, respectively, and $N(t)$ and $X(t)$ are their lapse functions.
The constant $k$ in both metrics (\ref{FRW-1}) and (\ref{FRW-2}) is the spatial curvature, whose positive, vanishing and negative values ($k=1,0,-1$) correspond to the closed, flat and open universes, respectively.

We solve the equations of motion (\ref{e.o.m.}), and obtain the following equations:
\begin{eqnarray}\label{Friedmann-1}
(ab\!-\! 1)\dot{A}(t) \!-\!\xi(t) \!~N(t)=0,
\end{eqnarray}
and
\begin{eqnarray}\label{Friedmann-2}
\xi(t)\!\!~\dot{A}(t) \!-\! a\!~\xi^{2}(t) X\!(t)
\!+\!\Big(\! \frac{ab\!-\! 1}{\ell^{2}}A^{2}(t)\! -\! k\!\Big) N\!(t)\!=\! 0,~~
\end{eqnarray}
together with a relation between two scale factors as:
\begin{eqnarray}
Y(t)=b~A(t),
\end{eqnarray}
and the following relations for the spin connection fields:
\begin{eqnarray}
\omega^{0}(t)=\frac{\sqrt{1\!-\! kr^{2}}}{ab\!-\!1} ~d\varphi,~~~~~~~~~~~~~~~~~~~~~~~~~~~~
\nonumber\\
\omega^{1}(t)=\frac{r\Big(kab-(ab\!-\!1)^{2}~\xi^{2}(t)\Big)}{(ab\!-\!1)^{2}~\xi(t)} ~d\varphi,~~~~~~~~
\nonumber\\
\omega^{2}(t)=\frac{\xi(t)}{\sqrt{1\!-\! kr^{2}}} ~dr,~~~~~~~~~~~~~~~~~~~~~~~~~~~~~
\nonumber\\
\tilde{\omega}^{0}(t)=\frac{-a\!~b^{2}\sqrt{1\!-\! kr^{2}}}{(ab\!-\!1)^{2}} ~d\varphi,~~~~~~~~~~~~~~~~~~~~~
\nonumber\\
\tilde{\omega}^{1}(t)=-\frac{br\Big(kab+\ell^{-2} (ab\!-\!1)^{3}\!~A^{2}(t)\Big)}{(ab\!-\!1)^{2}~\xi(t)} ~d\varphi,
\nonumber\\
\tilde{\omega}^{2}(t)=\frac{b\!~\ell^{-2} (ab\!-\!1)~A^{2}(t)}{\xi(t)\sqrt{1\!-\! kr^{2}}} ~dr,~~~~~~~~~~~~~~~~~
\end{eqnarray}
where $b$ is an arbitrary constant, dot denotes the time derivative ($\dot{A}\equiv\frac{dA}{dt}$), and
\begin{eqnarray}
\xi(t)\!=\!\sqrt{\!-k\!-\!\ell^{-2} (ab\!-\!1)^{2}A^{2}(t)},~~~~ |A(t)|<\frac{\ell\sqrt{\!-k}}{|ab\!-\!1|},
\end{eqnarray}
which implies that we have an open universe with negative spatial curvature ($k=-1$), where the radial coordinate $r$ is defined on $0\leq r<+\infty$.
Solving the equations (\ref{Friedmann-1}) and (\ref{Friedmann-2}) give the following relations for $N(t)$ and $X(t)$ in terms of the scale factor $A(t)$:
\begin{eqnarray}
N(t)=\frac{ab\!-\!1}{\xi(t)}\dot{A}(t),
\end{eqnarray}
and
\begin{eqnarray}
X(t)=-\frac{bk}{\xi^{3}(t)}\dot{A}(t).
\end{eqnarray}
The equations (\ref{Friedmann-1}) and (\ref{Friedmann-2}) do not restrict the scale factor $A(t)$ of the FRW metric (\ref{FRW-1}), and then $A(t)$ is an arbitrary function of the timelike coordinate $t$.
Using the following coordinate transformation
\begin{eqnarray}
\hat{t}\equiv \ell ~ arcsin\Big(\frac{ab\!-\! 1}{\ell}A(t)\Big),
\end{eqnarray}
the FRW metric (\ref{FRW-1}) can be rewritten as:
\begin{eqnarray}\label{FRW-3}
ds^{2}=-d\hat{t}^{2}+\hat{a}^{2}(\hat{t}) \Big(\frac{dr^{2}}{1\!+\! r^2}+r^{2}d\varphi^{2}\Big),
\end{eqnarray}
where the scale factor is
\begin{eqnarray}
\hat{a}(\hat{t})=\frac{\ell\!~ sin(\hat{t}/\ell)}{ab-1},
\end{eqnarray}
which is obviously an oscillating solution.
The Hubble parameter for this solution is obtained as follows:
\begin{eqnarray}
H(\hat{t})\equiv\frac{\dot{\hat{a}}}{\hat{a}}=\frac{1}{\ell} cot(\hat{t}/\ell).
\end{eqnarray}
Its deceleration parameter is
\begin{eqnarray}
q(\hat{t})\equiv-\frac{a\ddot{\hat{a}}}{\dot{\hat{a}}^{2}}= tan^{2}(\hat{t}/\ell),
\end{eqnarray}
which is obviously positive and implies that the expansion of the universe is decelerating.
Using another coordinate transformation as follows:
\begin{eqnarray}
t' \equiv \frac{bk A(t)}{\sqrt{1\!-\!\ell^{-2} (ab\!-\!1)^{2}A^{2}(t)}},
\end{eqnarray}
the second FRW metric (\ref{FRW-2}) can be rewritten in the following form:
\begin{eqnarray}\label{FRW-4}
df^{2}=-dt'^{2}+\bar{a}^{2}(t') \Big(\frac{dr^{2}}{1\!+\! r^2}+r^{2}d\varphi^{2}\Big),
\end{eqnarray}
where the scale factor is
\begin{eqnarray}
\bar{a}(t')=\frac{b\!~ t'}{\sqrt{b^{2}+\ell^{-2} (ab\!-\!1)^{2}\!~ t'^{2}}},
\end{eqnarray}
The Hubble and deceleration parameters for this solution are
\begin{eqnarray}
H(t')=\frac{b^{2}}{t\Big(b^{2}+\ell^{-2} (ab\!-\!1)^{2}\!~ t'^{2}\Big)},
\end{eqnarray}
and
\begin{eqnarray}\label{deceleration parameter}
q(t')=\frac{3(ab\!-\!1)^{2}\!~ t'^{2}}{\ell^{2}b^{2}},
\end{eqnarray}
respectively. The positive deceleration parameter (\ref{deceleration parameter}) implies that the universe which is described by the FRW metric (\ref{FRW-4}) has a decelerating expansion.

\section{Conclusions}

We have obtained a Lie bialgebra for the AdS Lie algebra in (2+1)-dimensional spacetime.
Applying a Manin triple corresponding to the AdS Lie bialgebra as a gauge symmetry algebra of the Chern-Simons theory, we have introduced a new (2+1)-dimensional bi-metric gravity model. Our ghost-free Chern-Simons bi-gravity action is an exactly soluble model without any local degrees of freedom which describes two interacting massless spin-2 fields. Its free field limit has also no local degrees of freedom, and is a ghost-free action. Our model is different from known three dimensional bi-metric massive gravity theories such as dRGT and ZDG models.
The black hole solution (\ref{BTZ metric})-(\ref{AdS spin connection}) of our model is different from the previously obtained three dimensional bi-gravity black hole solutions \cite{2Banados,Afshar}. In our solution, one of the metrics is the BTZ black hole metric, and the other metric is a new black hole metric with two curvature singularities and two horizons unlike two coordinate singularities of the previously obtained solutions of the
three dimensional $f\!-g$ theory. Our solution is also different from the other solutions in its asymptotic behaviour, and unlike other solutions, it has not asymptotically AdS form.
Our bi-gravity model admits a homogeneous and isotropic FRW cosmological solution with two different scale factors in two metrics $g_{\mu\nu}$ and $f_{\mu\nu}$, which describe two  universes with the decelerating expansions.
It is also interesting to study details of the new black hole metric (\ref{New metric}) as well as gravity/CFT correspondence at the boundary of the bi-gravity model (\ref{model}), which we leave them to later.
Chern-Simons formulation of our interacting model simplifies its quantization, which may be interesting in the context of quantum gravity.
Study of (3+1)-dimensional version of the AdS Lie bialgebra, and resultant (3+1)-dimensional gauge invariant interacting model is also a useful task which may have interesting features.

\textbf{Acknowledgments:}
We would like to express our heartfelt gratitude to M.M. Sheikh-Jabbari, F. Darabi, M.R. Setare and F. Loran for their useful comments and discussions.
This research was supported by a research fund No. 217D4310 from
Azarbaijan Shahid Madani university.






\begin{thebibliography}{99}

\bibitem{1E.Witten} {E. Witten, Nucl. Phys. \textbf{B311} (1988/89) 46-78.}
\bibitem{1S.Carlip} {S. Carlip, J. Korean Phys. Soc. \textbf{28} (1995) S447, [arXiv:gr-qc/9503024].}
\bibitem{2S.Carlip} {S. Carlip, Living Rev. Rel. \textbf{8} (2005) 1.}
\bibitem{3S.Carlip} {S. Carlip, Class. Quant. Grav. \textbf{22} (2005) R85-R124.}
\bibitem{2E.Witten} {E. Witten, [arXiv:0706.3359[hep-th]].}
\bibitem{1Boulanger} {N. Boulanger, T. Damour, L. Gualtieri and M. Henneaux, Nucl.Phys. \textbf{B597} (2001) 127-171.}
\bibitem{2Boulanger} {N. Boulanger and L. Gualtieri, Class. Quant. Grav. \textbf{18} (2001) 1485.}
%
\bibitem{1K.Hinterbichler} {K. Hinterbichler, Rev. Mod. Phys. \textbf{84} (2012) 671-710.}
\bibitem{1C.de Rham} {C. de Rham, Living Rev. Relativity \textbf{17} (2014) 7.}
\bibitem{1S.Deser} {S. Deser, R. Jackiw and S. Templeton, Annals Phys. \textbf{140} (1982) 372-411.}
\bibitem{2S.Deser} {S. Deser, R. Jackiw and S. Templeton, Phys. Rev. Lett. \textbf{48} (1982)
975-978.}
\bibitem{4S.Carlip} {S. Carlip, JHEP \textbf{0810} (2008) 078.}
\bibitem{1M.Blagojevic} {M. Blagojevic and B. Cvetkovic, JHEP \textbf{0905} (2009) 073.}
\bibitem{1E.A.Bergshoeff} {E. A. Bergshoeff, O. Hohm and P. K. Townsend, Phys. Rev. Lett. \textbf{102} (2009) 201301.}
\bibitem{2E.A.Bergshoeff} {E. A. Bergshoeff, O. Hohm and P. K. Townsend, Phys. Rev. \textbf{D79} (2009) 124042.}
\bibitem{2M.Blagojevic} {M. Blagojevic and B. Cvetkovic, JHEP \textbf{1101} (2011) 082.}
\bibitem{O.Hohm} {O. Hohm, A. Routh, P. K. Townsend and B. Zhang, Phys. Rev. \textbf{D86}
(2012) 084035.}
\bibitem{2C.de Rham} {C. de Rham and G. Gabadadze, Phys. Rev. \textbf{D82} (2010) 044020.}
\bibitem{3C.de Rham} {C. de Rham, G. Gabadadze and A. J. Tolley, Phys. Rev. Lett. \textbf{106} (2011) 231101.}
\bibitem{1S.F.Hassan} {S.F. Hassan and R.A. Rosen, JHEP \textbf{1107} (2011) 009.}
\bibitem{W.Merbis} {W. Merbis, PhD thesis defended at the University of Groningen, 2014.}
\bibitem{3E.A.Bergshoeff} {E.A. Bergshoeff, S. de Haan, O. Hohm, W. Merbis and P.K. Townsend, Phys. Rev. Lett. \textbf{111} (2013) 111102.}
\bibitem{4E.A.Bergshoeff} {E.A. Bergshoeff, A.F. Goya, W. Merbis and J. Rosseel, JHEP \textbf{1404} (2014) 012.}
\bibitem{2S.F.Hassan} {S.F. Hassan and R.A. Rosen, JHEP \textbf{1202} (2012) 126.}
\bibitem{2K.Hinterbichler} {K. Hinterbichler and R.A. Rosen, JHEP \textbf{1207} (2012) 047.}

\bibitem{5E.A.Bergshoeff} {E. Bergshoeff, O. Hohm,W. Merbis, A.J. Routh and P.K. Townsend, Lect. Notes Phys. \textbf{892} (2015) 181-201.}

\bibitem{M.R.Setare} {M.R. Setare, H. Adami, Phys. Lett. \textbf{B750} (2015) 31-36.}
\bibitem{H.Bacry} {H. Bacry, P. Combe and J. L. Richard, Nuovo Cim. {\bf A67}, (1970) 267-299 ; ibid. {\bf A70}, 289-312 (1970).}
\bibitem{R.Schrader} R. Schrader, Fortsch. Phys. {\bf 20} (1972) 701-734.
\bibitem{1J.A.de Azcarraga} {J.A. de Azcarraga, K. Kamimura and J. Lukierski, Phys. Rev. \textbf{D83}, (2011) 124036.}
\bibitem{2J.A.de Azcarraga} {J.A. de Azcarraga, K. Kamimura and J. Lukierski, Int. J. Mod. Phys. Conf. Ser. \textbf{23} (2013) 01160.}
\bibitem{P.Salgado} {P. Salgado, R.J. Szabo, O. Valdivia, Phys. Rev. \textbf{D89}, (2014) 084077.}
\bibitem{1O.Cebecioglu} {O. Cebecio\u{g}lu, S. Kibaro\u{g}lu, Phys. Rev. \textbf{D90} (2014) 084053.}
\bibitem{2O.Cebecioglu} {O. Cebecio\u{g}lu, S. Kibaro\u{g}lu, Phys. Lett. \textbf{B751} (2015) 131-134.}
\bibitem{1S.Hoseinzadeh} {S. Hoseinzadeh and A. Rezaei-Aghdam, Phys. Rev. \textbf{D90} (2014) 084008.}
\bibitem{D.V.Soroka} {D.V. Soroka and V.A. Soroka, Phys. Lett. \textbf{B707} (2012) 160-162.}
\bibitem{J.Diaz} {J. D\'{i}az, O. Fierro, F. Izaurieta, N. Merino, E. Rodr�guez, P. Salgado, O. Valdivia, J. Phys. \textbf{A45} (2012) 255207.}
\bibitem{2S.Hoseinzadeh} {S. Hoseinzadeh and A. Rezaei-Aghdam, Eur. Phys. J. \textbf{C75} (2015) 227.}
%
\bibitem{3S.Hoseinzadeh} {S. Hoseinzadeh and A. Rezaei-Aghdam, [arXiv:1705.11042[hep-th]].}
\bibitem{V.Chari} {V. Chari and A. Pressley, Cambridge Univ. Press. 1994.}
\bibitem{A.Rezaei-Aghdam} {A. Rezaei-Aghdam, M. Hemmati, A.R. Rastkar, J. Phys. A: Math. Gen. \textbf{38} (2005) 3981-3994.}
\bibitem{C.Nappi} {C.R. Nappi and E. Witten, Phys. Rev. Lett. \textbf{71}, (1993) 3751-3753.}
\bibitem{1Banados} {M. Ba\~{n}ados, C. Teitelboim and J. Zanelli, Phys. Rev. Lett. \textbf{69} (1992) 1849; M. Ba\~{n}ados, M. Henneaux, C. Teitelboim and J. Zanelli, Phys. Rev. \textbf{D48} (1993) 1506.}
\bibitem{2Banados} {M. Ba\~{n}ados and S. Theisen, JHEP \textbf{0911} (2009) 033.}
\bibitem{Afshar} {H.R. Afshar, M. Alishahiha and A. Naseh, Phys. Rev. \textbf{D81} (2010) 044029.}

\end{thebibliography}
\end{document}